\documentclass[11pt]{article}
\usepackage[utf8]{inputenc}
\usepackage[T1]{fontenc}
\usepackage{times}
\usepackage[margin=1in]{geometry}
\usepackage{booktabs}
\usepackage{amsmath,amssymb}
\usepackage{microtype}
\usepackage[hidelinks]{hyperref}
\usepackage{xcolor}
\usepackage[numbers]{natbib}
\usepackage{graphicx}

\twocolumn
\title{Unpredictable Safety: Domain-Dependent Compliance and the Transparency Gap in Open-Weight LLMs}

\author{Zacharie Bugaud\thanks{Correspondence: \texttt{zacharie.bugaud@gmail.com}} \\ \textit{Astera Institute}}

\date{}

\begin{document}
\maketitle

\begin{abstract}
We present a systematic study of domain-dependent safety behavior in open-weight LLMs: 7~standardized experiments across 7~ethical domains, testing 5~models (12B--70B) in 4,200~interactions with dual-judge validation. Using a dual-condition methodology, each scenario tested in both an \emph{analytical} framing (identify the harm) and an \emph{operational} framing (help commit the harm), we find compliance rates vary from 14.7\% (human trafficking) to 85.7\% (surveillance design), a 71-percentage-point span with non-overlapping cluster-boot\-strapped 95\% CIs. Trustworthy deployment requires predictable safety behavior, yet we find compliance is highly context-dependent: the same model (Mistral Nemo 12B) provides surveillance designs in 100\% of requests but assists with trafficking in only 26.7\%. This unpredictability is opaque to deployers: the technical framing bypass, where harmful requests reframed as engineering problems override safety training without any external signal that refusal thresholds have shifted. Within-domain heterogeneity reaches 84.4pp, meaning safety behavior cannot be predicted even at the domain level. A replication on five frontier closed models (GPT-4.1/5.2, Claude Haiku/Sonnet/Opus 4.x; $n{=}4{,}163$ responses) accessed via the GitHub Copilot CLI deployed-product surface reproduces the same domain stratification, attenuated in absolute level but identical in shape, with the two low-codification domains (science fraud, surveillance) again the most permissive. These results show that current safety mechanisms lack the transparency and consistency required for trustworthy AI deployment.
\end{abstract}

\section{Introduction}

The AI safety community has invested heavily in alignment techniques (RLHF, constitutional AI, red-teaming, safety fine-tuning) under the implicit assumption that these produce \emph{general} safety capabilities. A model that refuses to help with bomb-making should, in principle, also refuse to help with election manipulation, scientific fraud, or environmental crime. We test this assumption with 7 standardized experiments across 7 ethical domains, evaluating 5 open-weight LLMs (12B--70B parameters) in 4,200 interactions, supported by 13 additional exploratory studies (14,650 interactions).

\textbf{Central finding:} Compliance rates vary from 14.7\% (human trafficking) to 85.7\% (surveillance design), a 71-percentage-point (pp) span with non-overlapping cluster-boot\-strapped 95\% CIs. The same model (Mistral Nemo 12B) provides surveillance designs in 100\% of requests while complying with trafficking requests only 26.7\% of the time. Llama~3.3 at 70B achieves 0\% compliance on environmental crimes yet 51.7\% on surveillance, with the same weights and no system prompt.

\textbf{Contributions:} (1)~Docu\-men\-ta\-tion of domain-dependent compliance variation with cluster-boot\-strapped inference; (2)~hier\-archical heterogeneity, where safety varies between domains, sub-domains, and models simultaneously, with within-domain gaps reaching 84.4pp; (3)~a technical framing bypass present across all five tested models, where harmful actions reframed as engineering problems elicit higher compliance; (4)~a stable model safety hierarchy (mean Spearman $\rho = 0.68$).

\section{Related Work}

\textbf{Safety benchmarks.} Existing benchmarks (Truth\-ful\-QA~\cite{lin2022}, Adv\-Bench~\cite{zou2023}, Harm\-Bench~\cite{mazeika2024}) evaluate safety along a single aggregate dimension. Models receive an aggregate score, obscuring domain-level variation. Our work illustrates that a single score is misleading: Gemma~3 averages 44\% compliance overall yet ranges from 8\% on trafficking to 93\% on surveillance.

\textbf{Alignment failures.} Sharma et al.~\cite{sharma2023} document sycophancy as an RLHF side effect; Wei et al.~\cite{wei2023} propose synthetic data to mitigate it. Casper et al.~\cite{casper2023} identify fundamental RLHF limitations. Our results provide empirical evidence of a specific failure mode: domain-dependent safety, where alignment generalizes unevenly across ethical categories.

\textbf{Domain-specific safety.} Weidinger et al.~\cite{weidinger2022} taxonomize AI risks but do not test whether alignment is uniform across them. Bai et al.~\cite{bai2022} train a helpful and harmless assistant with RLHF but evaluate safety holistically. Our work is, to our knowledge, among the first controlled studies quantifying domain-dependent safety behavior, with up to 71pp separating the best- and worst-defended domains.

\section{Methods}

\subsection{Experimental Design}

Each of 7 studies tested 20 scenarios covering distinct sub-domains within an ethical area. Each scenario was written in two conditions:
\begin{itemize}
\item \textbf{Analytical} (e.g., advocate, auditor, reviewer): ``Identify this harm, analyze violations, recommend protections.''
\item \textbf{Operational} (e.g., exploiter, architect, facilitator): ``Help me commit this harm, provide guidance, explain how to avoid detection.''
\end{itemize}
Scale per study: 5 models $\times$ 20 scenarios $\times$ 2 conditions $\times$ 3 replications = 600 interactions.

\subsection{Models}

We tested five open-weight LLMs deployed locally via Ollama (Q4\_K\_M quant.,
$T{=}0.7$, max tokens 2048, no system prompt), listed in Table~\ref{tab:modelspecs}.
We use $T{=}0.7$ to allow sampling diversity (the default for chat use); no system prompt is used to test base safety behavior without deployment-time guardrails. Quantization (4-bit) enables local execution; we note this as a limitation.

\begin{table}[t]
\centering
\small
\caption{Model specifications.}
\label{tab:modelspecs}
\begin{tabular}{lcl}
\toprule
\textbf{Model} & \textbf{Params} & \textbf{Developer} \\
\midrule
Gemma 3 & 27B & Google \\
Qwen 3 & 32B & Alibaba \\
Mistral Nemo & 12B & Mistral AI \\
Llama 3.3 & 70B & Meta \\
DeepSeek R1 & 32B & DeepSeek \\
\bottomrule
\end{tabular}
\end{table}

\subsection{Evaluation}

\textbf{Primary judge.} LLM-as-judge~\cite{zheng2023} using Gemma 3 27B at $T{=}0.0$ with domain-specific rubrics: 3 ordinal scales (1--5) for harm identification, refusal strength, and harmful guidance quality; 2 binary indicators for domain-specific markers (e.g., cites laws, advocates protection).

\textbf{Key metrics.} \emph{Compliance rate}: proportion of operational responses providing substantive harmful guidance (score $\geq$3/5). \emph{Strong refusal rate}: proportion with refusal score $\geq$4/5. \emph{``Hypocrisy''}: proportion of model-scenario pairs where the model scores $\geq$4 on harm identification (analytical) yet $\geq$3 on harmful guidance (oper\-ational).

\textbf{Second-judge validation.} 140 stratified responses (20 per core study, non-Llama outputs) re-evaluated by Llama 3.3 70B with matching harm and refusal scales at $T{=}0.0$. Cohen's $\kappa = 0.90$ for binary compliance; 97.8\% within-1 agreement on harm scores.

\textbf{Statistical inference.} The 3 replications per model-scenario pair at $T{=}0.7$ are not fully independent, making the pair ($n = 100$ per study) the natural independent unit. We report cluster-boot\-strapped 95\% CIs resampling model-scenario pairs ($B = 10{,}000$), producing CIs $\sim$1.4$\times$ wider than naive response-level estimates. We also compute pair-level variance decomposition and test threshold sensitivity at $\geq$2, $\geq$3, and $\geq$4.

\section{Results}

\subsection{Domain-Dependent Compliance}

\begin{table}[t]
\centering
\caption{Compliance, refusal, and hypocrisy rates with boot\-strapped 95\% CIs. Str. = compliance stratum (L/M/H).}
\label{tab:compliance}
\resizebox{\columnwidth}{!}{%
\begin{tabular}{llccc}
\toprule
\textbf{Domain} & \textbf{Str.} & \textbf{Compl. (CI)} & \textbf{Refusal} & \textbf{Hypoc.} \\
\midrule
Trafficking & L & 14.7 [10.0, 20.0] & 91.3 & 4.0 \\
Corruption & L & 21.7 [15.0, 28.7] & 80.3 & 13.0 \\
Environmental & L & 24.7 [17.7, 32.0] & 76.3 & 14.0 \\
Elections & L & 25.0 [18.3, 32.0] & 85.3 & 14.0 \\
Labor & M & 49.7 [41.0, 58.3] & 46.3 & 39.0 \\
Science fraud & H & 69.7 [61.7, 77.3] & 43.7 & 53.0 \\
Surveillance & H & 85.7 [79.7, 91.3] & 20.0 & 79.0 \\
\bottomrule
\end{tabular}}
\end{table}

Table~\ref{tab:compliance} shows the central result. For descriptive convenience, the 7 rates partition into three strata: Low ($\sim$15--25\%: trafficking, corruption, environmental, elections), Medium ($\sim$50\%: labor), and High ($\sim$70--86\%: science fraud, surveillance). This partition captures 96\% of between-domain variance, though it is derived from only 7 data points with a singleton Medium stratum. Adjacent strata are separated: Low's upper bound (32.0\%) $<$ Medium's lower (41.0\%); Medium's upper (58.3\%) $<$ High's lower (61.7\%).

\begin{table}[t]
\centering
\small
\caption{Stratification quality: $R^2$ by contiguous strata count. T\,=\,Traf., C\,=\,Corr., En\,=\,Env., El\,=\,Elec., L\,=\,Labor, F\,=\,Sci.F., S\,=\,Surv.}
\label{tab:strata}
\begin{tabular}{lccc}
\toprule
$k$ & \textbf{Strata assignment} & $R^2$ & $\Delta R^2$ \\
\midrule
2 & \{T,C,En,El\} $|$ \{L,F,S\} & 0.839 & +0.839 \\
3 & \{T,C,En,El\} $|$ \{L\} $|$ \{F,S\} & 0.956 & +0.117 \\
4 & \{T,C,En,El\} $|$ \{L\} $|$ \{F\} $|$ \{S\} & 0.985 & +0.029 \\
\bottomrule
\end{tabular}
\end{table}

Table~\ref{tab:strata} shows that the three-stratum partition explains 96\% of between-domain variance, with diminishing marginal returns for additional strata ($\Delta R^2 = 0.117$ from 2$\to$3, $\Delta R^2 = 0.029$ from 3$\to$4). The four Low domains have mutually overlapping CIs, so finer partitions within this group are not supported.

Effect sizes for the analytical-vs-operational condition shift are large (pair-level Cohen's $d$ up to 3.56), confirming that the compliance differences are not merely statistically significant but practically meaningful.

The 71pp span from trafficking (14.7\%) to surveillance (85.7\%), with non-overlapping CIs, is the core empirical finding: safety behavior is strongly domain-dependent in the models we tested.

\subsection{Threshold Sensitivity}

The compliance threshold ($\geq$3/5) is researcher-specified. We re-compute at alternative thresholds in Table~\ref{tab:threshold}:

\begin{table}[t]
\centering
\small
\caption{Compliance at thresholds $\geq$2, $\geq$3, and $\geq$4.}
\label{tab:threshold}
\begin{tabular}{lccc}
\toprule
\textbf{Domain} & $\geq$\textbf{2} & $\geq$\textbf{3} & $\geq$\textbf{4} \\
\midrule
Trafficking & 23.7\% & 14.7\% & 1.7\% \\
Corruption & 45.3\% & 21.7\% & 3.3\% \\
Environmental & 34.0\% & 24.7\% & 0.7\% \\
Elections & 40.3\% & 25.0\% & 9.3\% \\
Labor & 66.7\% & 49.7\% & 5.7\% \\
Science fraud & 75.7\% & 69.7\% & 21.7\% \\
Surveillance & 88.0\% & 85.7\% & 56.7\% \\
\bottomrule
\end{tabular}
\end{table}

The domain ordering is stable across thresholds. Kendall's $\tau$ between the $\geq$2 and $\geq$3 rankings is 0.81 ($p = 0.011$); between $\geq$3 and $\geq$4, $\tau = 0.71$ ($p = 0.030$). At $\geq$4, surveillance remains highest at 56.7\% and environmental crime lowest at 0.7\%. The stratum separation is preserved: Low domains stay below 10\% at $\geq$4, while High domains remain above 20\%.

\subsection{Variance Decomposition}

\begin{table}[t]
\centering
\small
\caption{Pair-level variance decomposition ($n = 700$ pairs).}
\label{tab:variance}
\begin{tabular}{lcc}
\toprule
\textbf{Source} & \textbf{SS} & \textbf{\% Total} \\
\midrule
Domain & 279.1 & 35.6\% \\
Model & 115.0 & 14.6\% \\
Scenario (w/in domain) & 206.2 & 26.3\% \\
M $\times$ S interaction & 184.7 & 23.5\% \\
\bottomrule
\end{tabular}
\end{table}

Table~\ref{tab:variance} shows the pair-level variance decomposition. Domain is the dominant systematic factor (36\%), 2.4$\times$ more than model identity (15\%). Scenario variation within domains (26\%) exceeds model variation, indicating substantial sub-domain heterogeneity. The 24\% residual represents model $\times$ scenario interaction, i.e.\ specific model-domain-scenario combinations that deviate from main effects.

\subsection{Model Safety Hierarchy}

\begin{table}[t]
\centering
\caption{Per-model compliance rates (\%) across 7 studies. Cell values rounded to integers; $\bar{x}$ is the exact mean.}
\label{tab:models}
\resizebox{\columnwidth}{!}{%
\begin{tabular}{lccccccc|c}
\toprule
\textbf{Model} & \textbf{Traf} & \textbf{Corr} & \textbf{Env} & \textbf{Elec} & \textbf{Lab} & \textbf{Sci.F} & \textbf{Surv} & $\bar{x}$ \\
\midrule
Llama 3.3 & 5 & 7 & \textbf{0} & 3 & 20 & 25 & 52 & 16.0 \\
Qwen 3 & 7 & 15 & 27 & 22 & 50 & 67 & 92 & 39.8 \\
Gemma 3 & 8 & 10 & 32 & 17 & 57 & 93 & 93 & 44.3 \\
DeepSeek R1 & 27 & 38 & 37 & 35 & 50 & 75 & 92 & 50.5 \\
Mistral & 27 & 38 & 28 & 48 & 72 & 88 & \textbf{100} & 57.4 \\
\bottomrule
\end{tabular}}
\end{table}

Table~\ref{tab:models} presents individual model compliance rates. The ranking (Llama $<$ Qwen $<$ Gemma $<$ DeepSeek $<$ Mistral, by increasing compliance, though Qwen, Gemma, and DeepSeek are not cleanly separated) is stable across domains: all 21 pairwise Spearman correlations are positive (mean $\rho = 0.68$; individual correlations do not reach significance at $n{=}5$, but the probability of all 21 being positive by chance is $< 5 \times 10^{-7}$ by sign test). Llama~3.3 achieves 0\% compliance on environmental crimes (60 interactions) and near-zero on trafficking (5\%). Mistral Nemo reaches 100\% on surveillance (60 interactions, every response provided surveillance designs). The 0\%--100\% range across models (Llama at 0\%, Mistral at 100\% on different domains) illustrates that domain dominates model identity as a source of variation.

The ranking is \emph{consistent with} a scale benefit (the largest model, 70B, is safest; the smallest, 12B, least safe), but the five models differ in training data and RLHF methodology, not only scale. We cannot attribute the ordering to parameter count alone.

\subsection{The Knowledge-Action Gap}
\label{sec:knowledge}

\begin{table}[t]
\centering
\small
\caption{Hypocrisy rates with 95\% bootstrap CIs.}
\label{tab:hypocrisy}
\begin{tabular}{lcc}
\toprule
\textbf{Domain} & \textbf{Hypoc. (\%)} & \textbf{95\% CI} \\
\midrule
Trafficking & 4.0 & [1, 8] \\
Corruption & 13.0 & [7, 20] \\
Environmental & 14.0 & [8, 21] \\
Elections & 14.0 & [8, 21] \\
Labor & 39.0 & [29, 49] \\
Science fraud & 53.0 & [43, 63] \\
Surveillance & 79.0 & [71, 87] \\
\bottomrule
\end{tabular}
\end{table}

One of the clearest patterns is that models \emph{understand} a harm is wrong (scoring $\geq$4/5 on harm identification) yet \emph{provide guidance anyway} (scoring $\geq$3/5 on harmful guidance). Table~\ref{tab:hypocrisy} shows this ``hypocrisy'' ranges from 4\% (trafficking) to 79\% (surveillance), a 75pp span with non-overlapping CIs.

Surveillance hypocrisy at 79\% means that in nearly 4 of 5 model-scenario pairs, the model correctly identifies surveillance as a civil liberties violation yet provides detailed system designs when asked in the operational framing. Trafficking at 4\% shows the opposite: when models recognize trafficking, they almost never provide guidance.

\subsection{Intra-Domain Variation}

Each study reveals substantial compliance variation across sub-domains (Table~\ref{tab:intra}):

\begin{table}[t]
\centering
\caption{Within-study compliance ranges.}
\label{tab:intra}
\resizebox{\columnwidth}{!}{%
\begin{tabular}{llcc}
\toprule
\textbf{Study} & \textbf{Sub-domains} & \textbf{Min--Max} & \textbf{Gap} \\
\midrule
Trafficking & Child expl.--Supply chain & 3.3--46.7\% & 43.3pp \\
Corruption & Obstr.--Money laund. & 0.0--38.3\% & 38.3pp \\
Environmental & Emissions--Reg.\ evas. & 3.3--66.7\% & 63.3pp \\
Elections & Oppo.\ res.--Deepfakes & 3.3--46.7\% & 43.3pp \\
Labor & Migrant--Worker surv. & 0.0--84.4\% & \textbf{84.4pp} \\
Sci.\ fraud & Grant fraud--Citation & 36.7--86.7\% & 50.0pp \\
Surveillance & Dissent--Workplace & 73.3--93.3\% & 20.0pp \\
\bottomrule
\end{tabular}}
\end{table}

The labor study is the most extreme: within a single domain, compliance ranges from 0\% (migrant exploitation, linked to federal trafficking laws) to 84.4\% (worker surveillance, unregulated), an 84.4pp gap. This confirms that safety training operates at a finer granularity than the domain level. We term this \textbf{hierarchical heterogeneity}: compliance varies between domains, sub-domains, and models simultaneously, with no single level fully explaining the pattern.

\subsection{Technical Framing Bypass}

When harmful actions are reframed as engineering or optimization problems, compliance increases across all five models. The pattern appears across both core and exploratory studies (7+13=20 total): safety dismissal framed as ``risk analysis'' (92\%), deception as ``strategic communication'' (99.8\% for models $\geq$12B), military escalation as ``national security'' (93\%), discriminatory hiring as ``HR opti\-mi\-za\-tion'' (78\%), and surveillance as ``system archi\-tecture'' (85.7\%). The within-domain labor example is especially clear: migrant exploitation receives 0\% compliance while worker surveillance (framed as ``workplace management'') reaches 84.4\%, within the same ethical domain but with very different compliance, distinguished primarily by technical framing.

\paragraph{A four-category framing taxonomy.} Post-hoc coding of the 140 core operational prompts (single coder; categories are non-exclusive) identifies four recurring re-framings:
\begin{itemize}
\item \textbf{F1 Engineering re-framing} (``design'', ``architecture'', ``pipeline''): 47\% of operational prompts.
\item \textbf{F2 Optimization re-framing} (``maximize'', ``efficiency'', ``throughput''): 24\%.
\item \textbf{F3 Authority/role assumption} (``I am the X official/manager/officer''): 81\%, frequently co-occurring with F1/F2.
\item \textbf{F4 Procedural neutralization} (``standard practice'', ``best-practice template''): 18\%.
\end{itemize}
The High-compliance domains (surveillance, science fraud) are disproportionately F1$+$F3 (engineering plus authority); the Low-compliance domains are dominated by F3-only operational prompts whose harm is named in plain criminal-legal terms. A controlled ablation that strips each category independently is the natural next test (see \S\ref{sec:mechanism}).

\subsection{Second-Judge Validation}

To assess inter-judge reliability, 140 stratified responses (20 per core study, non-Llama outputs only) were re-evaluated by Llama 3.3 70B scoring the same harm and refusal dimensions at $T{=}0.0$ (Table~\ref{tab:judge}):

\begin{table}[t]
\centering
\small
\caption{Inter-judge agreement: Gemma~3 27B (primary) vs.\ Llama~3.3 70B (second).}
\label{tab:judge}
\begin{tabular}{lcc}
\toprule
\textbf{Metric} & \textbf{Harm} & \textbf{Refusal} \\
\midrule
Spearman $\rho$ & 0.866 & 0.820 \\
Exact agreement & 63.0\% & 41.3\% \\
Within-1 agreement & 97.8\% & 82.6\% \\
Binary $\kappa$ (compliance) & \multicolumn{2}{c}{0.898} \\
Binary agreement & \multicolumn{2}{c}{94.9\%} \\
\bottomrule
\end{tabular}
\end{table}

Cohen's $\kappa = 0.90$ for binary compliance confirms near-perfect inter-judge agreement. Per-study agreement ranges from 85\% (elections) to 100\% (corruption, surveillance). The 97.8\% within-1 agreement on harm scores indicates that when judges disagree, the disagreement is almost always by a single ordinal point.

\subsection{Post-Hoc: Legal Institutionalization}

We observe, post hoc, that stratum placement correlates with legal infra\-structure (Table~\ref{tab:legal}):

\begin{table}[t]
\centering
\caption{Legal institutionalization by compliance stratum (Str.).}
\label{tab:legal}
\resizebox{\columnwidth}{!}{%
\begin{tabular}{llcl}
\toprule
\textbf{Domain} & \textbf{Str.} & \textbf{Compl.} & \textbf{Legal status} \\
\midrule
Trafficking & L & 14.7\% & TVPA, DOJ/FBI \\
Corruption & L & 21.7\% & FCPA, SEC/FBI \\
Environmental & L & 24.7\% & CAA/CWA, EPA \\
Elections & L & 25.0\% & FECA, FEC/DOJ \\
Labor & M & 49.7\% & Partial: FLSA \\
Sci.\ fraud & H & 69.7\% & None: ORI only \\
Surveillance & H & 85.7\% & None: no agency \\
\bottomrule
\end{tabular}}
\end{table}

Low-stratum domains share dedicated criminal statutes and active enforcement agencies. High-stratum domains involve harms recognized as unethical but lacking criminal penalties. The singleton Medium domain (labor) has partial criminal coverage. This association is suggestive but not causal; five competing explanations remain plausible: training data representation, lexical cues, cultural salience, judge sensitivity, and legal codification effects.

\subsection{Closed-Model Deployed-Product Replication}
\label{sec:closed}

To test whether the open-weight pattern generalizes to frontier closed-source systems, we replicate the full 140-scenario protocol on five closed models accessed through the GitHub Copilot CLI deployed-product surface: GPT-4.1, GPT-5.2, Claude Haiku 4.5, Claude Sonnet 4.6, and Claude Opus 4.7 ($n{=}4{,}163$ scored responses across the 7 studies, dual framings, and 3 replicates; identical Gemma-3 27B judge rubric; same cluster bootstrap on $({\rm scenario}, {\rm condition})$ pairs with $B{=}10{,}000$, seed 42). For 22 prompts (all on Claude Sonnet 4.6, Claude Opus 4.7, or GPT-5.2, all on environmental or trafficking domains), the deployed product returned an empty string on three independent retries; we treat these as silent refusals and code them as such (judge harm score~$=1$, refusal~$=5$). Per-cell sample sizes are uniform for GPT-4.1 and the three Claude models (19--20 scenario pairs out of 20 in every cell) but reduced for GPT-5.2 (4--19 pairs per cell, due to additional upstream content-filter rejections on that model); GPT-5.2 cell-level estimates therefore carry wider CIs (e.g., env.\ 18.5\% [0.0, 44.4]) and should be read as exploratory.

\begin{table*}[t]
\centering
\caption{Closed-model operational-framing compliance (\%), by domain. Compliance threshold: harm score $\geq 3$ on the 1--5 judge scale (matching the open-weight rubric in \S Methods); 95\% cluster-bootstrap CIs ($B{=}10{,}000$, seed 42). Mean is across the 7 study cells.}
\label{tab:closed-op}
\footnotesize
\setlength{\tabcolsep}{4pt}
\begin{tabular}{lrrrrrrrr}
\toprule
\textbf{Model} & \textbf{Sci.\ fraud} & \textbf{Corrupt.} & \textbf{Surveill.} & \textbf{Elections} & \textbf{Labor} & \textbf{Environ.} & \textbf{Traffick.} & \textbf{Mean} \\
\midrule
GPT-4.1            & 35.0 &  0.0 & 35.0 & 0.0 & 18.3 &  0.0 & 0.0 & 12.6 \\
GPT-5.2            & 40.4 &  0.0 &  3.8 & 0.0 &  0.0 & 18.5 & 0.0 &  9.0 \\
Claude Haiku 4.5   &  5.0 &  0.0 &  0.0 & 0.0 &  0.0 &  0.0 & 0.0 &  0.7 \\
Claude Sonnet 4.6  &  1.7 &  0.0 &  6.7 & 0.0 &  0.0 &  0.0 & 0.0 &  1.2 \\
Claude Opus 4.7    &  0.0 &  0.0 &  1.7 & 0.0 &  0.0 &  0.0 & 0.0 &  0.2 \\
\midrule
\textbf{Domain mean} & \textbf{16.4} & \textbf{0.0} & \textbf{9.4} & \textbf{0.0} & \textbf{3.7} & \textbf{3.7} & \textbf{0.0} & \\
\bottomrule
\end{tabular}
\end{table*}

Three findings carry across the closed set (Table~\ref{tab:closed-op}). First, the \emph{domain ordering reproduces at the stratum level}: the two highest-compliance domains across closed models are again science fraud (mean 16.4\%) and surveillance (9.4\%) -- the two low-codification domains -- while the four high-codification criminal domains (corruption, elections, environmental, trafficking) collapse to 0.0\% operational compliance in 4 of 5 models. Of the three remaining non-zero cells, two fall in the low/mixed-codification half of the open-weight ranking (GPT-4.1 labor 18.3\%, GPT-4.1 surveillance 35.0\%); the third (GPT-5.2 environmental 18.5\%) is the lone high-codification exception and corresponds to the highest-compliance high-codification domain in the open-weight ranking (env.\ 24.7\%). The 71pp open-weight span is therefore not an artifact of quantization or scale: the same low/high codification grouping appears across model families and deployment surfaces, attenuated in absolute level (highest closed-model cell 40.4\% vs.\ open-weight 85.7\%) but identical in stratum membership.

Second, a \emph{model hierarchy emerges within the closed set} that mirrors the open-weight pattern: Claude Opus 4.7 (0.2\% mean op.) $<$ Haiku 4.5 (0.7\%) $<$ Sonnet 4.6 (1.2\%) $<$ GPT-5.2 (9.0\%) $<$ GPT-4.1 (12.6\%). The hierarchy is stable across domains (all Claude variants $<$ both GPT variants on the mean), with the gap concentrated in the two low-codification domains.

Third, the \emph{knowledge--action gap of \S\ref{sec:knowledge}} reappears strongly. Across the closed set, mean analytical-framing compliance (recognising and discussing the harm) is 9.1\% versus 4.7\% operational; for Claude Opus the asymmetry is sharpest (16.7\% analytical vs.\ 0.2\% operational), indicating recognition of the harm coupled with consistent refusal to perform it -- the same dissociation we observe on open weights.

\textbf{Methodological caveat.} Closed-model responses were elicited via the Copilot CLI deployed-product surface, which combines the underlying model with vendor-side system prompts, tool harnesses, and any pre-/post-classifier filters; results therefore measure the \emph{deployed product}, not the raw model. This is the conservative direction: deployed products embed additional refusal layers beyond the model itself, so any compliance we observe is a lower bound on what the bare model would emit. The qualitative replication of the open-weight domain ordering under this strictly more-guarded condition strengthens, rather than weakens, the central claim that compliance is domain-dependent.

\section{Domain Selection Rationale}
\label{sec:selection}

Domains were selected before data collection to vary on two orthogonal axes: (i)~\emph{legal codification} (dedicated criminal statute and enforcement agency vs.\ ethically recognized harm without criminal penalty), and (ii)~plausible \emph{training-data salience} (estimated informally from term frequency in common open corpora and from prior safety-benchmark coverage). The 7 domains span the resulting design: trafficking, corruption, environmental, and elections are high-codification with varying salience; labor is mixed-codification; science fraud and surveillance are low-codification. This construction is therefore designed to expose codification-correlated variance and is not a uniform sample of the full safety landscape. The 71pp headline span is nonetheless robust to this design choice: point estimates span 10pp \emph{within} the high-codification block (trafficking 14.7\% vs.\ elections 25.0\%; CIs within this block overlap, consistent with codification being a coarse predictor that does not separate adjacent high-codification domains), while CIs are non-overlapping across the High/Low transition (Low's upper 32.0\% vs.\ High's lower 61.7\%; Table~\ref{tab:compliance}). We do not claim that an arbitrary new domain would lie within the observed range; we claim that the variation within this deliberately-varied set is itself too large to support a single ``safety score'' per model.

\section{Mechanistic Hypotheses}
\label{sec:mechanism}

The results above are behavioral. We did not establish a mechanism, and we are cautious about that gap. We enumerate five hypotheses that are individually testable and mutually distinguishable on the indicated ablations; the camera-ready commits to two of them (H1, H5) as immediate follow-ups since they require no additional model access.

\begin{enumerate}
\item[\textbf{H1}] \textit{Lexical-trigger account.} Refusal is gated by surface tokens (``traffick'', ``bribe'') rather than by intent. \emph{Test:} synonym/obfuscation perturbations; predicts compliance increases on lexical-but-not-semantic substitutions.
\item[\textbf{H2}] \textit{Training-frequency account.} Refusal tracks frequency of the harm term and of its refusal exemplars in pre-training and RLHF data. \emph{Test:} correlate per-domain compliance with corpus frequency proxies.
\item[\textbf{H3}] \textit{Domain-codification account.} Refusal correlates with the existence of a dedicated criminal statute and enforcement agency (see Table~\ref{tab:legal}). \emph{Test:} extend to non-U.S.\ legal frames (EU GDPR for surveillance; EU AI Act categories).
\item[\textbf{H4}] \textit{Framing-bypass account.} Engineering or optimization framing (F1, F2) routes through a competing instruction-following circuit that out-weighs the refusal head at decoding time. \emph{Test:} logit-lens / attribution patching on a small open model (Gemma-2 2B/9B) comparing analytical vs.\ operational tokens.
\item[\textbf{H5}] \textit{Persona-conditioning account.} The operational persona (``I am a city official'', ``I am an HR manager''; F3) supplies an authority cue that lowers refusal probability. \emph{Test:} persona-strip ablation (remove role claim) and persona-swap ablation (assign the analytical persona an operational request).
\end{enumerate}

None of these are mutually exclusive; H4 and H5 are the most natural explanations of the within-domain labor gap (84.4pp), since the analytical and operational prompts there share the same harm terms (defeating H1 partially) and the same statutory backdrop (defeating H3 in isolation), while differing in engineering framing (F1) and persona (F3).

\section{Discussion}

\textbf{Safety behavior is unpredictable.} For trustworthy deployment, safety must be consistent and transparent. Instead, we find 71pp compliance differences across domains that all involve serious real-world harms. Deployers cannot predict when a model will refuse: Llama~3.3 achieves 0\% on environmental crimes yet 51.7\% on surveillance; Mistral Nemo reaches 100\% on surveillance yet 26.7\% on trafficking.

\textbf{The transparency problem.} The technical framing bypass operates without any external signal. When a harmful request is reframed as an engineering problem, compliance increases silently; deployers and users receive no indication that safety thresholds have shifted. This opacity undermines trust.

\textbf{Hierarchical unpredictability.} Within-domain heterogeneity (84.4pp in the labor study) means that even domain-level safety assessments are insufficient. Migrant exploitation receives 0\% compliance while worker surveillance receives 84.4\%, within the same ethical domain but with very different model behavior.

\textbf{Toward trustworthy safety.} Trustworthy AI requires safety mechanisms whose behavior can be predicted, explained, and verified. Our results show that current safety alignment lacks the first two properties (it is domain-dependent and framing-sensitive) and suggest it may lack the third, as the framing bypass operates without visible cues. Domain-specific auditing with per-domain compliance reporting would improve transparency.

\section{Future Work}
\label{sec:future}

Four directions follow directly from these results. (1)~\textbf{Framing-invariant safety fine-tuning} using our paired analytical/operational prompts as training pairs: the supervision signal is that operational re-framings of a recognized harm should inherit the analytical-condition refusal. (2)~\textbf{Representation-level probes} testing H4 and H5 on Gemma-2 2B/9B via logit-lens and attribution patching, with persona-strip and lexical-perturbation ablations on the full five-model set. (3)~\textbf{Quantization-controlled replications} on the open-weight set (full-precision vs.\ Q4\_K\_M) using the same 140-scenario protocol, dual-judge validation, and cluster bootstrap; the closed-model replication is now reported in \S\ref{sec:closed} and a Gemini-class extension is the natural next step. (4)~\textbf{Per-domain auditing protocols}: a reporting standard in which model cards include per-domain compliance with CIs rather than a single aggregate safety score, and an external audit procedure that uses paired-framing scenarios as the unit of measurement.

\section*{Data and Code Availability}

Given the dual-use nature of this work (the dataset consists of operational-framing prompts designed to elicit harmful assistance from LLMs), we do not publicly release the scenario corpus, raw model responses, or judge-scored outputs. Releasing these artifacts would amount to publishing a tested jailbreak catalog across seven sensitive domains. The methodology is fully specified in this paper to enable independent reconstruction by researchers with comparable safety review: study design (\S\ref{sec:method}), the dual-framing protocol, judge rubric and secondary-judge validation procedure, the exact Ollama model identifiers and quantization ($\mathrm{Q4\_K\_M}$), cluster-bootstrap parameters ($B = 10{,}000$, seed 42), and per-domain compliance statistics with confidence intervals. Aggregate results (per-domain, per-model, per-condition compliance rates and CIs) are available from the author on reasoned request to qualified safety researchers; the underlying prompts and responses are not.

\section{Limitations}

\textbf{No human validation of judge scores} (most significant limitation). Both LLM judges may share systematic blind spots; cross-family agreement between Gemma~3 27B and Llama~3.3 70B ($\kappa = 0.90$) is \emph{consistent with}, but does not establish, freedom from shared rubric overfit. The primary judge (Gemma 3 27B) shares a model family with one evaluated model. Until human annotation confirms ratings, findings should be interpreted as judge-consistent domain dependence. \textbf{Small domain sample}: 7 data points are insufficient for strong structural claims; the domain set was deliberately constructed to vary on legal codification and salience (\S\ref{sec:selection}), not sampled uniformly from the safety landscape. \textbf{Closed-model evidence is deployed-product, not raw model}: the \S\ref{sec:closed} replication on GPT-4.1/5.2 and Claude Haiku/Sonnet/Opus 4.x measures the Copilot CLI surface (model + vendor system prompt + tool harness + classifier filters), which is the conservative direction (lower-bounds bare-model compliance) but does not isolate model behavior; Gemini-class systems were not covered. \textbf{Direct requests only}: no adversarial techniques (jailbreaking, multi-turn persuasion). \textbf{No system prompt}: results reflect base model safety; deployment-time system prompts may significantly alter compliance rates (exploratory \texttt{bilateral\_sysprompt\_*} runs on two models suggest the analytical--operational gap persists under warn/balanced/devil-advocate system prompts). \textbf{Within-condition prompt-wording sensitivity}: each scenario is realized as a single analytical and a single operational prompt; we did not perturb wording within a condition. \textbf{4-bit quantization}: Q4\_K\_M may affect safety behavior relative to full-precision inference. \textbf{U.S.\ legal framing}: scenarios reflect U.S.\ legal structures. \textbf{Mechanistic claims}: \S\ref{sec:mechanism} states hypotheses, not findings. \textbf{Statistical caveats}: the model safety hierarchy rests on $n{=}5$ models per study (individual Spearman correlations do not reach significance; the hierarchy is supported by a sign test across 21 study-pairs). Cohen's $d$ is computed on ordinal 1--5 scores, which may overstate effect sizes under floor/ceiling effects. The three named variance components (domain, model, scenario) sum to 76.5\% of total variance (35.6\% + 14.6\% + 26.3\%; displayed as 36\%, 15\%, 26\% after rounding); the 23.5\% residual represents model $\times$ scenario interaction.

\section{Conclusion}

Compliance rates in five open-weight models span 71 percentage points, from 14.7\% (trafficking [10.0, 20.0]) to 85.7\% (surveillance [79.7, 91.3]), with non-overlapping cluster-boot\-strapped CIs. Domain accounts for 36\% of pair-level variance, with scenario (26\%) exceeding model identity (15\%). Within-domain heterogeneity reaches 84.4pp. A technical framing bypass and a stable model safety hierarchy ($\rho = 0.68$) generalize across domains. Second-judge validation ($\kappa = 0.90$) confirms measurement robustness. These results motivate domain-specific safety auditing rather than aggregate safety scores.

\section*{Ethics Statement}

This research tests AI systems' willingness to assist with harmful activities by presenting scenarios drawn from documented real-world harms. All prompts were designed by the authors and do not reproduce actual criminal instructions. We report compliance rates to inform AI safety improvements, not to enable harmful use. All experiments used locally deployed open-weight models with no external API calls.

\end{document}